\documentclass[a4paper,12pt]{article}

\begin{document}

\title{Deterministic Local Conversion of Incomparable States by Collective LOCC}

\author{Indrani Chattopadhyay
\thanks{indrani@cucc.ernet.in} ,
 and Debasis Sarkar
\thanks{dsarkar@cucc.ernet.in, dsappmath@caluniv.ac.in}}

\maketitle
\begin{center}
Department of Applied Mathematics, University of Calcutta,

92, A.P.C. Road, Kolkata- 700009, India

\end{center}

\begin{abstract}

 Incomparability of pure bipartite entangled states under deterministic
LOCC is a very strange phenomena. We find two possible ways of
getting our desired pure entangled state which is incomparable with
the given input state, by collective LOCC with certainty. The first
one is by providing some pure entanglement through the lower
dimensional maximally-entangled states or using further less amount
of entanglement and the next one is by collective operation on two
pairs which are individually incomparable. It is quite surprising
that we are able to achieve maximally entangled states of any
Schmidt rank from a finite number of $2\times 2$ pure entangled
states only by deterministic LOCC. We provide general theory for the
case of $3\times3$ system of incomparable states by the above
processes where incomparability seems to be the most hardest one.

 PACS number(s): 03.67.Hk, 03.65.Bz, 03.67. -a, 89.70 +c.

\end{abstract}
\maketitle

\vspace{1cm}

\section{Introduction}
The necessity of quantum entanglement on different quantum
information and communication tasks ~\cite{mes,cryp} naturally
raises the question of manipulating entanglement shared between
different parties. It is very often useful when shared parties are
able to manipulate entanglement by local operation and classical
communication(LOCC). Within this restricted scenario, Bennet et
al. ~\cite{locc} provided the asymptotic conversion of
entanglement by LOCC for bipartite systems. Things are
qualitatively different if we restrict ourselves to finite regime,
i.e., only finite number of copies are available, even for pure
bipartite entangled states. Nielsen ~\cite{nielsen} provided a
necessary and sufficient condition for the local conversion of
pure bipartite entangled states for single copy case with
certainty. Vidal extended this to the case of probabilistic local
conversion (SLOCC) of two pure bipartite entangled states
~\cite{vidal}. Further, Morikoshi ~\cite{morikoshi} investigated
the recovery of entanglement loss in the process of local
conversion and several other groups studied the possibility and
impossibility of entanglement manipulation in different
context~\cite{plenio,som1,mutual
cat,cat1,som2,som3,leung,trump,Duan1,incom}.

In many practical situation it is often necessary to use a
particular entangled state for a specific task. Therefore the
possibility of extending the set of states which produces the target
state of our interest, beyond those satisfying Nielsen's criterion,
is of importance. Jonathan and Plenio~\cite{plenio} investigated the
case where in place of single copy transformation we also allow
collective operation by assigning some entanglement. The phenomena
is known as catalytic conversion. Similar to the situation of
partial recovery, Bandyopadhyay et.al.~\cite{som1,som2} and Feng
et.al.~\cite{mutual cat} studied the possibility of local conversion
of incomparable states (those not satisfying Nielsen's criteria) by
collective operation where some entanglement may be recovered. The
class of incomparable states  may be further reduced by the process
of multiple copy transformation~\cite{som3}. Still there remain
large class of incomparable states where deterministic conversion by
LOCC are not possible. Recently, Ishizaka~\cite{Ishizaka} showed
that using ppt-bound entanglement local conversion of any two states
(no matter what the Schmidt rank of the states are) is always
possible, at least with some probability. So the problem of finding
target state of our interest now remains for the case of
deterministic local conversion. In this paper our aim is to resolve
the incomparability of two pure bipartite states having the same
Schmidt rank, by deterministic LOCC. We first show that, using the
entanglement of some lower dimensional pure state, it is always
possible to break the incomparability in any dimension. This method
is found to be successful in providing maximally entangled states of
any dimension from next lower dimensional maximally entangled states
with the help of suitable non-maximally pure entangled states.
Actually it is possible to achieve any maximally entangled states by
a finite number of $2\times 2$ pure entangled states only. We also
observed that a special kind of mutual catalysis which we rather
call mutual co-operation is very much useful in converting the pairs
of incomparable states. We discuss exhaustively all our analysis for
$3\times 3$ systems in which incomparability seems to be the hardest
one.

\section{Background}

Let's take a quick overview of the topic of state transformation
by LOCC. We first mention Nielsen's criterion~\cite{nielsen} for
deterministic local conversion of $|\psi\rangle$ to
$|\phi\rangle,$ both are pure bipartite states shared between two
parties, say, Alice and Bob. Suppose up to local unitary
equivalence we write $|\psi\rangle$, $|\phi\rangle$ in the Schmidt
basis $\{|i_A\rangle ,|i_B\rangle \}$ with decreasing order of
Schmidt coefficients as : $|\psi\rangle= \sum_{i=1}^{d}
\sqrt{\alpha_{i}} |i_A i_B\rangle$, $|\phi\rangle= \sum_{i=1}^{d}
\sqrt{\beta_{i}} |i_A i_B\rangle,$ where $\alpha_{i}\geq
\alpha_{i+1}\geq 0$ and $\beta_{i}\geq \beta_{i+1}\geq0,$ for
$i=1,2,\cdots,d-1,$ $\sum_{i=1}^{d} \alpha_{i} = 1 =
\sum_{i=1}^{d} \beta_{i}$ and denote
$\lambda_\psi\equiv(\alpha_1,\alpha_2,\cdots,\alpha_d),$
$\lambda_\phi\equiv(\beta_1,\beta_2,\cdots,\beta_d)$. Then
Nielsen's criterion tells us that $|\psi\rangle\rightarrow|
\phi\rangle$ is possible with certainty under LOCC if and only if
$\lambda_\psi$ is majorized by $\lambda_\phi$ (denoted by
$\lambda_\psi\prec\lambda_\phi$); i.e.,
$\sum_{i=1}^{k}\alpha_{i}\leq \sum_{i=1}^{k}\beta_{i}$ for each
$k=1,2,\cdots,d$. One consequence of Nielsen's result is; if
$|\psi\rangle\rightarrow |\phi\rangle$ under LOCC with certainty,
then $E(|\psi\rangle)\geq E(|\phi\rangle)$ [where $E(\cdot)$
denote the von-Neumann entropy of the reduced density operator of
any subsystem and known as the entropy of entanglement]. If the
above criterion does not hold, then it is usually denoted by
$\mid\psi\rangle\not\rightarrow \mid\phi\rangle$. Though it may
happen that $|\phi\rangle\rightarrow |\psi\rangle$ under LOCC. Now
if it happens that $|\psi\rangle\not\rightarrow |\phi\rangle$ and
$|\phi\rangle\not\rightarrow |\psi\rangle$ then we denote it as
$|\psi\rangle\not\leftrightarrow |\phi\rangle$ and call
$\mid\psi\rangle, \mid\phi\rangle$ as a pair of incomparable
states. For $2\times2$ states it is always be the case that either
$|\psi\rangle\rightarrow |\phi\rangle$ or $|\phi\rangle\rightarrow
|\psi\rangle.$ Therefore, we look beyond the $2\times2$ system of
states. Now it is natural to ask that for such incomparable pairs,
is it not possible to convert $|\psi\rangle$ to $ |\phi\rangle$ by
means of LOCC? If we require $|\phi\rangle$ and we have a finite
but sufficient number of copies of $|\psi\rangle$ then Vidal's
theorem{\cite{vidal}} provide us the way of converting
probabilistically $|\psi\rangle$ to $|\phi\rangle$ by LOCC. This
is of course of no use for deterministic conversion. For the
purpose of transferring $|\psi\rangle$ to $|\phi\rangle$ where
$|\psi\rangle\not\rightarrow |\phi\rangle$ Jonathan and
Plenio~\cite{plenio} found that sometimes collective operation may
be useful to convert deterministically. They showed that if we
assist the conversion by another pure entangled state
$|\chi\rangle$, say catalyst, then the conversion $|\psi\rangle
\otimes |\chi\rangle \longrightarrow
|\phi\rangle\otimes|\chi\rangle$ may be possible by collective
LOCC deterministically. It is interesting in this method that we
recover all the entanglement used in the process. But what type of
pairs are really catalyzable? It is really hard to categorize.
Jonathan and Plenio~\cite{plenio} showed that if the conversion
$|\psi\rangle\rightarrow |\phi\rangle$ is possible by a catalytic
state then $\alpha_1 \leq \beta_1$ and $\alpha_d \geq \beta_d$
hold simultaneously. Also if $|\psi\rangle\rightarrow
|\phi\rangle$ is possible by catalysis, then $E(|\psi\rangle)\geq
E(|\phi\rangle)$. For $3\times3$ system of states, violation of
Nielsen's criteria always implies violation of the necessary
condition for catalysis. The existence of catalytic state is first
seen for $4\times4$ incomparable pairs and only in this level a
necessary and sufficient condition for the existence of $2 \times
2$ system of catalytic state is found until now
~\cite{Anspach,Duan}. Therefore except of some numerical evidence
it is really hard to find catalyzable pairs. Investigation in this
direction is going on by several groups~\cite{trump}. Now another
interesting result is provided by Feng et.al., and other
groups~\cite{mutual cat,som2} which is known as mutual catalysis.
The basic objective in this process is: given two pairs of
incomparable states, say, $|\psi_1\rangle\not\rightarrow
|\phi_1\rangle, |\psi_2\rangle\not\rightarrow |\phi_2\rangle,$
whether $|\psi_1\rangle \otimes |\psi_2\rangle
\longrightarrow|\phi_1\rangle\otimes|\phi_2\rangle$ is possible
under LOCC with certainty or not. Emphasis is given on the special
kind of mutual catalysis (in~\cite{som2} it is defined as super
catalysis), where in the conversion we recover not only the
entanglement assisted in the process but more than that, i.e., for
some incomparable pair $\mid\psi\rangle\not\rightarrow
\mid\phi\rangle$ there exists $(\mid\chi\rangle, \mid\eta\rangle)$
with $E(|\eta\rangle)\geq E(|\chi\rangle)$ and
$\mid\eta\rangle\rightarrow \mid\chi\rangle$ such that by
collective local operation $|\psi\rangle \otimes |\chi\rangle
\longrightarrow|\phi\rangle\otimes|\eta\rangle$ is possible
deterministically. It is interesting that the necessary condition
for the existence of such special kind of mutual catalytic pair
$(\mid\chi\rangle, \mid\eta\rangle)$, is the same as that for
catalyst. Hence this type of mutual catalysis is not possible for
$3\times3$ system of incomparability. It is shown, not
analytically, but by some numerical examples, that there are
systems for which catalyst does not exit but mutual catalysis
works. Trivially it is always possible that $|\psi\rangle \otimes
|\phi\rangle \longrightarrow|\phi\rangle\otimes|\psi\rangle$ under
LOCC with certainty. So existence of mutual catalytic state is
always possible. But it is not of use, as our target state
$|\phi\rangle$ is not in our hand and in the process of trivial
mutual catalysis we have to use it. Next to that, it is found by
Bandyopadhyay et.al.~\cite{som3},  sometimes if we increase the
number of copies of the states, then deterministic conversion of
incomparable states under LOCC may be possible; i.e.,
$\mid\psi\rangle\not\rightarrow \mid\phi\rangle$ but
$\mid\psi\rangle^{\otimes k}\rightarrow {\mid\phi\rangle}^
{\otimes k}$ is possible for some integer k. This phenomena is
called multi-copy transformation. A sufficient condition for an
incomparable pair to remain incomparable even if we increase the
number of copies as large as possible is that, either $\alpha_1 <
\beta_1$ and $\alpha_d < \beta_d$, or, $\alpha_1 > \beta_1$ and
$\alpha_d > \beta_d$ must hold simultaneously~\cite{som3}. We call
them as strongly incomparable~\cite{som3}. All pure incomparable
states in $3\times3$ are strongly incomparable. So all the process
of catalysis, mutual catalysis with some recovery and increasing
number of copies will fail for all $3\times3$ pure incomparable
pair of states and also in all pure strongly incomparable
bipartite classes. Therefore one may ask, is it not possible to
get the target state under LOCC for such incomparable states?

Here comes the question of using entanglement to reach the target
state. By the use of entanglement we mean to forget about
recovering the entanglement used in the process, but to
concentrate on converting the input state to the desired one. In
this paper we discover two paths from $|\psi\rangle$ to
$|\phi\rangle$ by collective LOCC deterministically, discussed in
two different sections. In the first part we show that for any
incomparable pair
$\mid\psi\rangle\not\leftrightarrow\mid\phi\rangle$, it is always
possible to locally transform the incomparable pair of states with
certainty, if we provide some amount of pure entanglement. We
shall show that by collective LOCC, on the joint system of
$|\psi\rangle$ and the next lower dimensional maximally entangled
state $|\Psi_{max}^{d-1}\rangle,$ we are always able to get the
target state $|\phi\rangle$ along with a product state
$|P\rangle.$ This indicates that in any finite dimension, an
incomparable pair can be made to transform if we have some supply
of pure entanglement, at most one copy of the next lower rank
maximally entangled state. Then it is only the matter that beside
of using $|\Psi_{max}^{d-1}\rangle,$ is it possible to use further
less amount of entanglement to achieve the target state, and if
yes, then what is the minimum amount of entanglement required for
such conversion. It is interesting to note here that in all the
pairs of incomparable states $|\psi\rangle$ and $|\phi\rangle$,
the first and last Schmidt coefficients of the states are
intricately related with each other. We complete our analysis of
minimum pure entanglement required for $3\times3$ system of
states, as it is the minimum and possibly the hardest dimension to
deal with. We also observe that the maximally entangled state of
any finite Schmidt rank together with a suitable pure entangled
state is able to produce the next higher rank maximally entangled
state under deterministic LOCC. Surprisingly we found that to
require a maximally entangled state of Schmidt rank $d$ we need
only $d-1$ number of suitable $2\times 2$ pure entangled states.

In the next section, we shall show another interesting phenomenon
that there may be two pairs of incomparable states such as
$\mid\psi\rangle\not\leftrightarrow \mid\phi\rangle$ and
$\mid\chi\rangle\not\leftrightarrow \mid\eta\rangle$ but
$|\psi\rangle \otimes |\chi\rangle
\longrightarrow|\phi\rangle\otimes|\eta\rangle$ is possible under
LOCC with certainty. That is if we require $|\phi\rangle$,
$|\eta\rangle$ but we have $|\psi\rangle$, $|\chi\rangle$ then
this collective operation may be possible. We call this phenomenon
as mutual co-operation. Obviously this is a general kind of mutual
catalysis which is the most preferred one. Beside of giving some
numerical evidences, we provide analytically an auxiliary pair of
incomparable states for every pair of incomparable states in
$3\times 3$ system such that the joint transformation under LOCC
is always possible with certainty. The concluding part of this
section reflects the feature of the $3\times3$ system that by
collective local operation on two copy of a state $|\psi\rangle$,
almost in all cases, we are able to get two different states
$|\phi_1\rangle$ and $|\phi_2\rangle$ both are incomparable with
$|\psi\rangle$. Now if we fix any one of $|\phi_1\rangle$ or
$|\phi_2\rangle$ as our target state, then we provide a good range
of the possible existence of the other state.

\section{Assistance by Entanglement}

Suppose we have a pair of  incomparable states $\mid\psi\rangle,
\mid\phi\rangle$ in $d\times d$ system, where the source state
$\mid\psi\rangle = \sum_{i=1}^{d} \sqrt{a_{i}}\mid ii\rangle,$ and
the target state $\mid\phi\rangle = \sum_{i=1}^{d} \sqrt{b_{i}}\mid
ii\rangle,$ are taken in their most general form with $a_{i} \geq
a_{i+1} \geq 0$ and $b_{i} \geq b_{i+1} \geq 0, \ \forall
i=1,2,\cdots,(d-1)$ together with $\sum_{i=1}^{d} a_{i}=
\sum_{i=1}^{d} b_{i} = 1$. In the whole processes we would not taken
into account the amount of entanglement of the states. Now, consider
the $(d-1)\times (d-1)$ maximally entangled state
$\mid\Psi_{max}^{d-1}\rangle =
\frac{1}{\sqrt{d-1}}\sum_{i=d+1}^{2d-1}\mid ii\rangle,$ and the
product state $\mid P\rangle = \mid 00\rangle. $ We want to make
possible the joint transformation $\mid\psi\rangle \otimes
\mid\Psi_{max}^{d-1}\rangle \rightarrow \mid\phi\rangle \otimes \mid
P\rangle$ under LOCC with certainty. For this we must have,
\begin{equation}
\frac{a_1k}{d-1}\leq\sum_{i=1}^{k}b_i; \forall \ k=1,2,\ldots,d-1.
\end{equation} To prove this, we first state a
theorem which shows an intricate relation holds between first and
last Schmidt coefficients of any two incomparable states.

{\bf Theorem.} For any pair of  incomparable states $\mid\psi\rangle
\not\leftrightarrow\mid\phi\rangle$ in $d\times d$ system, where
$\mid\psi\rangle = \sum_{i=1}^{d} \sqrt{a_{i}} \ \mid ii\rangle,$
and $\mid\phi\rangle = \sum_{i=1}^{d} \sqrt{b_{i}}\mid ii\rangle,$
with $a_{i} \geq a_{i+1} \geq 0, \sum_{i=1}^{d} a_{i} = 1$ and
$b_{i} \geq b_{i+1} \geq 0, \sum_{i=1}^{d} b_{i} = 1,$ the following
always holds:
\begin{equation}
a_1+b_d<1, \ b_1+a_d<1.
\end{equation}
Proof of this theorem follows from Nielsen's criteria if we analyze
incomparability condition critically.

The theorem above readily implies that $a_1< \sum_{i=1}^{d-1}
b_{i},$ from which we have
\begin{equation}
\frac{k{a_1}}{(d-1)}< \frac{k}{d-1} \sum_{i=1}^{d-1} b_{i} <
\sum_{i=1}^{k} b_{i}, \ \forall k=1,2,\cdots,{d-1}.
\end{equation}
So the incomparability condition itself implies that the required
joint transformation is possible under LOCC with certainty. i.e.,
$\mid\phi\rangle \otimes \mid\Psi_{max}^{d-1}\rangle \rightarrow
\mid\psi\rangle \otimes \mid P\rangle$ is possible under LOCC with
certainty. Therefore for any pair of incomparable states with a
given Schmidt rank the maximally entangled state of the next lower
rank is sufficient to assist the joint transformation under LOCC.

Next we show that instead of using lower rank maximally entangled
state, the conversion may be possible under LOCC if we use lower
rank non-maximally entangled states so that we need as much as
minimum use of the resource. We found explicitly the minimum amount
of entanglement that is required for the local transformation of any
$3\times 3$  incomparable pairs. Suppose $\mid\psi\rangle
,\mid\phi\rangle$ be a pair of  incomparable states such that
$\mid\psi\rangle = \sum_{i=1}^{3} \sqrt{a_{i}}\mid ii\rangle $ and
$\mid\phi\rangle = \sum_{i=1}^{3} \sqrt{b_{i}}\mid ii\rangle,$ where
$a_{1} \geq a_{2} \geq a_{3} \geq 0, \sum_{i=1}^{3} a_{i} = 1$ and
$b_{1} \geq b_{2} \geq b_{3} \geq 0, \sum_{i=1}^{3} b_{i} = 1$; and
consider a $2\times 2$ pure entangled state $\mid\chi\rangle =
\sqrt{c}\mid 44\rangle + \sqrt{1-c}\mid 55 \rangle, 1 > c \geq
\frac{1}{2},$ and a $2\times 2$ product state $\mid\eta\rangle =
\mid 44\rangle.$ Then the collective operation under LOCC
$\mid\psi\rangle \otimes \mid\chi\rangle \rightarrow \mid\phi\rangle
\otimes \mid\eta\rangle,$ occurs with certainty if $\mid\chi\rangle$
is specified according with the amount of entanglement used in the
process; i.e., $E(\chi)= - c\log_{2} c -(1-c)\log_{2} (1-c).$ Hence
to minimize E, we have to find the largest possible value of c,
obviously which is not 1. Now, we discuss separately two different
classes of $3\times 3$  incomparable pair of states.

{\bf{Type-1:}} When $a_{1} < b_{1}, a_{1}+ a_{2} > b_{1} + b_{2},$
then we must have $c \leq \frac{b_{1}+b_{2}}{a_{1}+a_{2}}$; i.e.,
the minimum amount of entanglement required in this process to
achieve $\mid\phi\rangle$ from $\mid\psi\rangle$ is $E=E_0$
corresponding to the value $c=c_0=
\frac{b_{1}+b_{1}}{a_{1}+a_{2}}.$

{\bf{Type-2:}} When $a_{1} > b_{1}, a_{1}+ a_{2} < b_{1} + b_{2},$
then we must have $c \leq \frac{b_{1}}{a_{1}};$ i.e., the minimum
amount of entanglement required in this process to achieve
$\mid\phi\rangle$ from $\mid\psi\rangle$ is $E=E_0$ corresponding
to the value $c=c_0= \frac{b_{1}}{a_{1}}.$

We conclude this section with an interesting result that in any
dimension  from a non-maximally pure entangled state
$\mid\psi^d\rangle$ of $d\times d$ system $(d \geq 3)$, we are able
to reach the maximally entangled state
$\mid\phi\rangle=\mid\Psi_{max}^d\rangle$ of the same dimension by
the use of the next lower dimensional maximally entangled state
$\mid\Psi_{max}^{d-1}\rangle$ through collective local operation
with certainty.

 {\bf Corollary-1.}  $\mid\psi^d\rangle \otimes
\mid\Psi_{max}^{d-1}\rangle \rightarrow \mid\Psi_{max}^{d}\rangle
\otimes \mid P\rangle,$ where $\mid P\rangle$ is a product state,
is possible under LOCC with certainty, if the largest Schmidt
coefficient, $a_1$ of $\mid\psi^d\rangle$ satisfies the relation
$a_1\leq\frac{d-1}{d}.$

 This result follows directly from the theorem above. In fact,
 instead of using the state $\mid\psi^d\rangle$, the
above transformation is possible by a $2\times 2$ state only.

{\bf Corollary-2.} The transformation $\mid\psi\rangle \otimes
\mid\Psi_{max}^{d-1}\rangle \rightarrow$ $ \mid\Psi_{max}^{d}\rangle
\otimes \mid P\rangle,$ is possible under LOCC with certainty, if we
take $\mid\psi\rangle$, as a $2\times 2$ state with Schmidt
coefficients $(\frac{d-1}{d}, \frac{1}{d}).$

Corollary-2  immediately suggests that it is possible to achieve a
maximally entangled state of any Schmidt rank $d, \ d \geq 3$ by
using a finite number of $2\times 2$ states only.

{\bf Corollary-3.} The transformation,

$\mid\psi_1\rangle \otimes \mid\psi_2\rangle \otimes
\cdots\mid\psi_{d-1}\rangle \rightarrow \mid\Psi_{max}^{d}\rangle
\otimes \mid P\rangle,$

\noindent is possible under LOCC with certainty, where
$\mid\psi_i\rangle, \ \forall i=1,2,\cdots,d-1$ are $2\times 2$
states with Schmidt coefficients $(\frac{d-i}{d-i+1},
\frac{1}{d-i+1})$, respectively.

\section{Mutual Co-operation}

In this section our main goal is to provide an auxiliary
incomparable pair so that the collective operation enables us to
find the desired states; i.e., given a pair  $\mid\psi\rangle
\not\leftrightarrow \mid\phi\rangle$ we want to find an auxiliary
pair $\mid\chi\rangle \not\leftrightarrow \mid\eta\rangle$ such that
$\mid\psi\rangle \otimes \mid\chi\rangle \rightarrow \mid\phi\rangle
\otimes \mid\eta\rangle,$ is possible under LOCC deterministically.
There are several ways to find nontrivial
$(\mid\chi\rangle,\mid\eta\rangle)$. We first provide some examples
that will show such features and then in two subsections we shall
give analytical results for $3\times 3$ system of incomparable
states. We explicitly provide the form of the auxiliary pair for all
possible incomparable pair $(\mid\psi\rangle,\mid\phi\rangle)$ in
$3\times 3$ system. One of the interesting feature of such
incomparable pairs is that we are unable to say that which state has
greater entanglement than the other. So in this way we may resolve
the incomparability of $(\mid\psi\rangle,\mid\phi\rangle)$ with
$E(\mid\psi\rangle)<E(\mid\phi\rangle)$ by mutual co-operation which
obviously claims that $E(\mid\chi\rangle)>E(\mid\eta\rangle)$. Other
interesting part we have studied analytically in $3\times 3$ system
is the following :

From two copy of a pure entangled state we are able to find two
different pure entangled states, both of which are incomparable
with the source state.
 Let us begin with an example of mutual co-operation.

\noindent{\em Example 1}.-- Consider a pair of pure entangled
states of the form

$$\mid\psi\rangle = \sqrt{0.4}\mid 00 \rangle +\sqrt{0.4}\mid 11
\rangle + \sqrt{0.2}\mid 22 \rangle,$$

$$\mid\phi\rangle = \sqrt{0.48}\mid 00 \rangle +\sqrt{0.26}\mid 11
\rangle + \sqrt{0.26}\mid 22 \rangle,$$

$$\mid\chi\rangle = \sqrt{0.49}\mid 33 \rangle +\sqrt{0.255}\mid 44
\rangle + \sqrt{0.255}\mid 55 \rangle,$$

$$\mid\eta\rangle = \sqrt{0.41}\mid 33 \rangle +\sqrt{0.41}\mid 44
\rangle + \sqrt{0.18}\mid 55 \rangle.$$

 It is easy to check that
 $\mid\psi\rangle \not\leftrightarrow
\mid\phi\rangle$ and $\mid\chi\rangle \not\leftrightarrow
\mid\eta\rangle$; whereas, $E(\mid\psi\rangle)\approx 1.5219
> E(\mid\phi\rangle) \approx 1.5188,  E(\mid\chi\rangle) \approx 1.5097 >
E(\mid\eta\rangle) \approx 1.5001$; and if we allow collective
operations locally on the joint systems, then the transformation
$\mid\psi\rangle \otimes\mid\chi\rangle\rightarrow\mid\phi\rangle
\otimes\mid\eta\rangle$ is possible with certainty, i.e., we see
that the two pairs which are  incomparable, will co-operate with
each other and make the joint transformation possible.

 If we looked upon the whole thing from a little more physical
point of view then something more comes out. We see here that the
comparability of the joint operation actually comes through the
co-operation with the comparable class, i.e., this four states are
related in such a way that $\mid\psi\rangle \rightarrow
\mid\eta\rangle$ and $\mid\phi\rangle \rightarrow
\mid\chi\rangle$. So here we reduce the incomparability of two
states by choosing some class of states comparable with them. It
is obvious that such a pair of states always exist for any
incomparable pair, i.e., incomparable pairs can always be made to
compare. Without going into details of the proof, we state that
this approach resolves the incomparability of the $3\times3$
states. At this moment someone may think that this result imply
that only with the help of comparable classes we destroy the
incomparability. Obviously, the answer is in the negative. The
next example is given in support of this.

\noindent{\em Example 2}.-- Consider two pairs of pure entangled
states $(\mid\psi\rangle,\mid\phi\rangle)$ and
$(\mid\chi\rangle,\mid\eta\rangle)$ of the form

$$\mid\psi\rangle = \sqrt{0.41}\mid 00 \rangle +\sqrt{0.38}\mid 11
\rangle + \sqrt{0.21}\mid 22 \rangle,$$

$$\mid\phi\rangle = \sqrt{0.4}\mid 00 \rangle +\sqrt{0.4}\mid 11
\rangle + \sqrt{0.2}\mid 22 \rangle,$$

$$\mid\chi\rangle = \sqrt{0.45}\mid 33 \rangle +\sqrt{0.34}\mid 44
\rangle + \sqrt{0.21}\mid 55 \rangle,$$

$$\mid\eta\rangle = \sqrt{0.48}\mid 33 \rangle +\sqrt{0.309}\mid 44
\rangle + \sqrt{0.211}\mid 55 \rangle.$$

It is quite surprising to see that not only $\mid\psi\rangle
\not\leftrightarrow \mid\phi\rangle$ and $\mid\chi\rangle
\not\leftrightarrow \mid\eta\rangle$ but also $\mid\psi\rangle
\not\leftrightarrow \mid\eta\rangle$, $\mid\chi\rangle
\not\leftrightarrow \mid\phi\rangle$. Beside this we also get the
extra facility to prepare $|\chi\rangle$ from $|\psi\rangle$ as
$\mid\psi\rangle \rightarrow \mid\chi\rangle$. From the
informative point of view the picture is although,
$E(\mid\psi\rangle)\approx 1.5307
> E(\mid\phi\rangle) \approx 1.5219,$ and  $E(\mid\chi\rangle) \approx
1.5204 > E(\mid\eta\rangle) \approx 1.50544,$ but still
independently we can not convert  $\mid\psi\rangle$ to either one of
$\mid\phi\rangle$ or $\mid\eta\rangle$ and also $\mid\chi\rangle$ to
either one of $\mid\phi\rangle$ or $\mid\eta\rangle$ with certainty
under LOCC. Therefore although the resource states have greater
information content, the individual pairs aren't convertible, but
treating them together we break their incomparability. Here we
didn't fix our eyes only on the transformation of the first pair and
recover as much as possible amount of entanglement from second pair
rather we have tried to reduce the  incomparability of both the two
pairs of states together.

To give rise the fact that mutual co-operation also exists in other
dimensions, we are providing other two sets of incomparable pairs in
$4 \times 4$ system which are strongly incomparable so that
deterministic local conversions are not possible by assisting also
catalytic states and $2 \times 2$ mutual catalytic states but
co-operate each to make the joint transformation possible.

\noindent{\em Example 3}.-- Consider two pairs of pure entangled
states $(\mid\psi\rangle,\mid\phi\rangle)$ and
$(\mid\chi\rangle,\mid\eta\rangle)$ of the form

$$\mid\psi\rangle = \sqrt{0.4}\mid 00 \rangle +\sqrt{0.3}\mid 11
\rangle + \sqrt{0.2}\mid 22 \rangle + \sqrt{0.1}\mid 33 \rangle,$$

$$\mid\phi\rangle = \sqrt{0.45}\mid 00 \rangle +\sqrt{0.29}\mid 11
\rangle + \sqrt{0.14}\mid 22 \rangle + \sqrt{0.12}\mid
33\rangle,$$

$$\mid\chi\rangle = \sqrt{0.5}\mid 44 \rangle +\sqrt{0.25}\mid 55
\rangle + \sqrt{0.2}\mid 66 \rangle + \sqrt{0.05}\mid 77
\rangle,$$

$$\mid\eta\rangle = \sqrt{0.48}\mid 44 \rangle +\sqrt{0.36}\mid 55
\rangle + \sqrt{0.12}\mid 66 \rangle + \sqrt{0.04}\mid
77\rangle.$$

It is easy to check that $\mid\psi\rangle \not\leftrightarrow
\mid\phi\rangle$ and $\mid\chi\rangle \not\leftrightarrow
\mid\eta\rangle,$ and $E(\mid\psi\rangle)\approx 1.846
> E(\mid\phi\rangle) \approx 1.800,$  $E(\mid\chi\rangle) \approx
1.680 > E(\mid\eta\rangle) \approx 1.592.$  However, one may check
$\mid\psi\rangle \otimes\mid\chi\rangle \longrightarrow \
\mid\phi\rangle \otimes\mid\eta\rangle,$ is possible under LOCC.

\noindent{\em Example 4}.-- Consider two pairs of pure entangled
states $(\mid\psi\rangle,\mid\phi\rangle)$ and
$(\mid\chi\rangle,\mid\eta\rangle)$ of the form

$$\mid\psi\rangle = \sqrt{0.4}\mid 00 \rangle +\sqrt{0.3}\mid 11
\rangle + \sqrt{0.2}\mid 22 \rangle + \sqrt{0.1}\mid 33 \rangle,$$

$$\mid\phi\rangle = \sqrt{0.45}\mid 00 \rangle +\sqrt{0.29}\mid 11
\rangle + \sqrt{0.14}\mid 22 \rangle + \sqrt{0.12}\mid
33\rangle,$$

$$\mid\chi\rangle = \sqrt{0.5}\mid 44 \rangle +\sqrt{0.23}\mid 55
\rangle + \sqrt{0.22}\mid 66 \rangle + \sqrt{0.05}\mid
77\rangle,$$

$$\mid\eta\rangle = \sqrt{0.48}\mid 44 \rangle +\sqrt{0.36}\mid 55
\rangle + \sqrt{0.12}\mid 66 \rangle + \sqrt{0.04}\mid
77\rangle.$$

Here also it is easy to verify that $\mid\psi\rangle
\not\leftrightarrow \mid\phi\rangle$, $\mid\psi\rangle
\not\leftrightarrow \mid\eta\rangle$, $\mid\chi\rangle
\not\leftrightarrow \mid\phi\rangle$ and $\mid\chi\rangle
\not\leftrightarrow \mid\eta\rangle.$ But surprisingly
$\mid\psi\rangle \rightarrow$   $\mid\chi\rangle$. Now it is very
interesting that we can prepare the state of co-operation from the
state in our hand. The relations between the entanglement of those
states are, $ E(\mid\psi\rangle) \approx 1.846 > E(\mid\phi\rangle)
\approx 1.800,$ and  $E(\mid\chi\rangle) \approx 1.684 >$
$E(\mid\eta\rangle) \approx 1.592,$  and $\mid\psi\rangle \otimes
\mid\chi\rangle \longrightarrow \mid\phi\rangle \otimes
\mid\eta\rangle$, is possible under LOCC with certainty. All the
examples we are providing are non-trivial one. Next we show some
analytical results for $3\times 3$ system of incomparable states.

\subsection{Local conversion of $3\times 3$ incomparable pairs by
auxiliary $3\times 3$ incomparable pairs}

Now we concentrate to the case of incomparable pairs in $3 \times 3$
system of states. We shall show for every pair of  incomparable pure
entangled states $(\mid\psi_1\rangle, \mid\phi_1\rangle)$ there is
always a pair of  incomparable pure entangled states
$(\mid\psi_2\rangle, \mid\phi_2\rangle)$ such that
$\mid\psi_1\rangle \ \otimes \mid\psi_2\rangle \longrightarrow \
\mid\phi_1\rangle \otimes$ $ \mid\phi_2\rangle,$ is possible under
LOCC with certainty. The main idea of this portion is, assuming
$\mid\psi_1\rangle$ as the source state and $\mid\phi_1\rangle$ as
the target state, we choose the nontrivial auxiliary incomparable
pair $(\mid\psi_2\rangle, \mid\phi_2\rangle)$ such that by
collective LOCC the joint transformation of both pairs is possible
with certainty.

Consider, $\mid\psi_1\rangle \equiv (a_1, a_2, a_3),
\mid\phi_1\rangle \equiv (b_1, b_2, b_3 )$ where $a_1 \geq a_2 \geq
a_3 \geq 0, a_1 + a_2 + a_3 =1, b_1 \geq b_2 \geq b_3 \geq 0, b_1 +
b_2 + b_3 =1.$ There are two possible cases of incomparability that
exist in this dimension, which are discussed and treated differently
below.

{\bf{Case-1:}}  $ a_1 > b_1, a_1+a_2 < b_1+b_2 .$ We choose
$\mid\psi_2\rangle \equiv (\beta_1, \beta_1,\beta_2),
\mid\phi_2\rangle \equiv (\alpha_1, \alpha_2, \alpha_2 )$ where
$\beta_1 > \beta_2 >0, 2\beta_ 1 +\beta_2 =1, \alpha_1 > \alpha_2
> 0, \alpha_1 +2 \alpha_2 =1, \beta_1 < \alpha_1, 2\beta_1 >
\alpha_1+\alpha_2,$ such that
\begin{equation}
\max\{\frac{a_1}{a_2}, \frac{b_1}{b_3}\}<\frac{\alpha_1}{\alpha_2}
\end{equation}
and
\begin{equation}
\frac{a_3}{(2a_1+a_3)}>\beta_2>\max\{\frac{\alpha_2b_3}{a_3},
\alpha_2(b_2+2b_3), \frac{\alpha_2(2-b_1)-a_3}{(1-a_3)},
1-\frac{\alpha_1 (b_1+b_2)}{a_1}\}
\end{equation}
Under such a choice the required joint transformation is always
possible.

In the above process there may arise a similar condition like our
first example. For this type of choice we have always
$\mid\psi_1\rangle\rightarrow \mid\phi_2\rangle.$ Except this choice
we further require that those cross pairs $(\mid\psi_1\rangle,
\mid\phi_2\rangle)$ or $(\mid\psi_2\rangle, \mid\phi_1\rangle),$
remain  incomparable too. To fulfill this requirement the state
$\mid\psi_2\rangle$ is chosen slight differently, as
$\mid\psi_2\rangle \equiv (\beta_1, \beta_2,\beta_3),$ where
$\beta_1 > \beta_2 > \beta_3 >0, \beta_ 1 +\beta_2 +\beta_3=1,$ such
that $a_1\beta_3>\beta_1 a_3>b_1\alpha_2,$ $a_1\beta_1<b_1\alpha_1,
a_3\beta_3
> b_3\alpha_2$ and $\{(\beta_1 a_3-a_2\beta_3)-(a_3-b_3)\}<
\min\{0, (\alpha_1 b_3-\alpha_2b_2), (a_2\beta_2-\alpha_2b_2) \}.$
After such a choice the pair $(\mid\psi_2\rangle,
\mid\phi_1\rangle)$ became incomparable except when $a_2=a_3.$ But
whenever we face the case $b_1=b_2$ and $a_2=a_3$ then
correspondingly we see that $\mid\psi_1\rangle\rightarrow
\mid\phi_2\rangle$ and $\mid\psi_2\rangle\rightarrow
\mid\phi_1\rangle.$

{\bf{Case-2:}} $a_1<b_1, a_1+a_2>b_1+b_2. $ In this case we choose
$\mid\psi_2\rangle \equiv (\beta_1, \beta_2,\beta_3),
\mid\phi_2\rangle \equiv (\alpha_1, \alpha_1, \alpha_2 )$ where
$\beta_1 > \beta_2 >\beta_3 >0, \beta_ 1 +\beta_2 +\beta_3=1,
\alpha_1 > \alpha_2 > 0, 2\alpha_1 + \alpha_2 =1, \beta_1 >
\alpha_1, \beta_1 +\beta_2 > 2\alpha_1.$ Now consider two
subcases.

Firstly, when $a_1< \frac{1}{2},$ we choose the state
$(\mid\psi_2\rangle ,\mid\phi_2\rangle)$  in such a way that
$\alpha_1 b_3
>a_1 \beta_3 >\beta_1 a_3$ and
\begin{equation}
\alpha_1>\max \{\frac{\beta_1 a_1}{b_1}, \frac{\beta_1 (a_1+a_2)+
a_1 \beta_2}{2b_1+b_2},\frac{ (1-\beta_3)(1-a_3)}{2(1-b_3)} \}
\end{equation}
Secondly, when $a_1 \geq \frac{1}{2},$ we choose the state
$(\mid\psi_2\rangle,$ $\mid\phi_2\rangle)$ in such a way that
$\beta_1 =\frac{1}{2}$ and $\alpha_1 b_3> a_1 \beta_3
>\beta_1 a_3,$
\begin{equation}
\alpha_1>\max \{\frac{a_1}{2 b_1}, \frac{a_1+a_2+ 2 a_1
\beta_2}{2(2b_1+b_2)}, \frac{(0.5+\beta_2)(1-a_3)}{2(1-b_3)},\frac{2
a_1+a_2}{4(b_1+b_2)}, \frac{a_1+a_2-a_2 \beta_3}{2-b_3} \}
\end{equation}
It is interesting to note that in the first subcase when $a_1= a_2$
then $\mid\psi_1\rangle\rightarrow \mid\phi_2\rangle.$ Except this,
our choice maintains $\mid\psi_i\rangle\not\leftrightarrow
\mid\phi_j\rangle, \ \forall i,j=1,2.$

\subsection{ Two incomparability with the same initial state may be broken jointly}

At the beginning of this section we want to present the special
result for $3\times 3$ system which is as follows :

For any source state $\mid\psi\rangle$ in $3\times 3$ system, with
distinct Schmidt coefficients there always exist two states
$(\mid\chi\rangle,$ $\mid\eta\rangle)$ such that both of them are
incomparable with $\mid\psi\rangle$ but from two copy of
$\mid\psi\rangle$ we are able to get them by collective LOCC with
certainty.

 Suppose the source state is $\mid\psi\rangle\equiv(a_1,
a_2, a_3)$ with $a_1> a_2> a_3>0, \ a_1+ a_2+ a_3 =1$ and the
other states are $\mid\chi\rangle\equiv(b_1, b_2, b_3)$ and
$\mid\eta\rangle\equiv(c_1, c_2, c_3)$ with $b_1> b_2> b_3>0$,
$b_1+ b_2+ b_3 =1$ and $c_1> c_2> c_3>0$, $ c_1+ c_2+ c_3 =1.$ We
first impose the incomparability conditions as $a_1>b_1,
a_1+a_2<b_1+b_2$ and $a_1< c_1, a_1+a_2 > c_1+c_2.$ Then it
follows from Nielsen's condition that there is always a possible
range of $(\mid\chi\rangle,\mid\eta\rangle)$ such that
$\mid\psi\rangle^{ \otimes 2} \longrightarrow
\mid\chi\rangle\otimes\mid\eta\rangle,$ under LOCC with certainty.
It should be noted that the cases of failure of this general
result is only the small number of cases where irrespective of the
incomparability condition, the Schmidt coefficients of the source
state are not all distinct, i.e., either $a_1=a_2$ or $a_2=a_3.$

This result is very important because we must keep in our mind the
fact, that multiple copy transformation is not possible for states
in $3 \times 3$ system. Now with this result in our hand, let us try
to fix $\mid\chi\rangle$ as our target state and find the possible
range (if exists at all) of $\mid\eta\rangle;$ i.e., in a quite
general sense we assume that there is two copy of the source state
$\mid\psi\rangle$ in our hand, where $\mid\psi\rangle
\not\leftrightarrow \mid\chi\rangle.$ Our aim is to find a
$\mid\eta\rangle;$ such that $\mid\psi\rangle \not\leftrightarrow
\mid\eta\rangle$, and to make possible the joint transformation
$\mid\psi\rangle ^{\otimes 2} \longrightarrow \mid\chi\rangle
\otimes\mid\eta\rangle,$ under LOCC with certainty.

Like the previous section here also we have two cases of
incomparability.

{\bf{Case-1:}}  When $a_1>b_1, a_1+a_2<b_1+b_2,$ we take
$\mid\eta\rangle \equiv (\alpha_1, \alpha_2, \alpha_2 )$ where
$\alpha_1 > \alpha_2> 0, \alpha_1 +2 \alpha_2 =1, a_1 < \alpha_1,
a_1+a_2 > \alpha_1+\alpha_2.$ Then the condition for such
transformation is, $a_3< \frac{1}{2}(1-\frac{{a_1}^2}{b_1}).$ Under
this condition we have not only one $\mid\eta\rangle,$ but a range
of it specified either by the relation
\begin{equation}
\alpha_2<\min\{\frac{a_1a_3}{b_1},
\frac{a_3^2}{b_3},\frac{a_3(2a_2+a_3)}{(b_2+2b_3)}\}, \ \mbox{for} \
{a_2}^2>a_1a_3,
\end{equation}

or by the relation
\begin{equation}
\alpha_2<\min \{ a_3+\frac{{a_2}^2-{a_3}^2}{2}, \frac{a_3^2}{b_3},
\frac{a_3(2a_2+a_3)}{(b_2+2b_3)} \}, \ \mbox{for} \ {a_2}^2<a_1a_3
\end{equation}

{\bf{Case-2:}} $a_1<b_1, a_1+a_2>b_1+b_2.$ Here we take
$\mid\eta\rangle \equiv (\alpha_1, \alpha_1, \alpha_2 )$ where
$\alpha_1 > \alpha_2 > 0, 2\alpha_1 + \alpha_2 =1, a_1 > \alpha_1,
a_1+a_2 < 2\alpha_1.$ Consider two subcases separately.

Firstly, when ${a_2}^2>a_1a_3$ then such a joint transformation
occurs if; $\frac{(a_1+a_2)^2}{2(b_1+b_2)}<a_1$ and the range of
$\mid\eta\rangle$ is specified by the relation
\begin{equation}
\alpha_1>\max\{ a_1-\frac{{a_1}^2-{a_2}^2}{2},
\frac{(a_1+a_2)^2}{2(b_1+b_2)},
\frac{(a_1)^2}{b_1},\frac{a_1(a_1+2a_2)}{(2b_1+b_2)}\}
\end{equation}
Next, when ${a_2}^2<a_1a_3$ then the condition for such
transformation is; $a_1+2a_2< 2b_1+b_2.$ Under this condition range
of $\mid\eta\rangle$ is specified by the relation
\begin{equation}
\alpha_1>\max\{a_1-\frac{{a_1}^2-{a_2}^2}{2},
\frac{a_1(2-a_1)}{(2-b_3)},\frac{(a_1)^2}{b_1},\frac{a_1(a_1+2a_2)}{(2b_1+b_2)}\}
\end{equation}
Finally we must mention that this process works for most of the
cases of  incomparability. But, it is not always successful; i.e.,
choosing any arbitrary incomparable pair we might not be able to
reach the target state by this method. This small range of failure
of the process is possibly due to the fact that we didn't ever
bother about the amount of entanglement contained into the states.
It is possible that $E(\mid\psi\rangle)\ll E(\mid\chi\rangle);$ for
which there doesn't exists such a state $\mid\eta\rangle,$
incomparable with $\mid\psi\rangle$ and $
E({\mid\psi\rangle}^{\otimes2})> E(\mid\chi\rangle \otimes
\mid\eta\rangle).$

In conclusion we have succeeded in providing a method by which any
incomparable pair of pure bipartite entangled states in any finite
dimension, can be maid to compare(i.e., transform one to another),
under LOCC with certainty, by providing some pure entanglement. We
observed that mutual co-operation is an useful process to break the
incomparability of two pairs under LOCC. This is not only discussed
as an abstract or rather complicated theory, but we provide the
algorithmic structure by which this goal can be really achieved.
This work supports the possibility  of reaching any pure bipartite
entangled states by deterministic LOCC.
 \vspace{1cm}

{\bf Acknowledgement.} We thank an anonymous referee for his
valuable comments and suggestions in revision of this paper. The
authors also thank S. Bandyopadhyay, G. Kar and S. Ghosh for useful
discussions in preparation of this paper. I.C. acknowledges CSIR,
India for providing fellowship during this work.

\end{document}